\documentclass[runningheads,natbib,onecollarge]{svjour2}
\smartqed  
\usepackage{graphicx}
\journalname{Space Science Review}
\begin{document}

\title{Asymmetries in Mars' Exosphere
}
\subtitle{Implications for X-ray and ENA Imaging}


\author{Mats Holmstr\"{o}m
}


\institute{Mats Holmstr\"{o}m \at
              Swedish Institute of Space Physics \\
              PO Box 812, SE-981 28 Kiruna, Sweden \\
              Tel.: +46-980-79186\\
              Fax: +46-980-79050\\
              \email{matsh@irf.se}           
}

\date{Received: April 5, 2006}

\maketitle

\begin{abstract}
Observations and simulations show that Mars' atmosphere has large seasonal 
variations. 
Total atmospheric density can have an order of magnitude latitudinal 
variation at exobase heights. 
By numerical simulations we show that these latitude variations 
in exobase parameters induce 
asymmetries in the hydrogen exosphere that propagate 
to large distances from the planet. 
We show that these asymmetries in the exosphere produce 
asymmetries in the fluxes of energetic neutral 
atoms (ENAs) and soft X-rays produced by charge exchange between 
the solar wind and exospheric hydrogen. 
This could be an explanation for asymmetries that have been 
observed in ENA and X-ray fluxes at Mars. 
\keywords{Mars \and Energetic Neutral Atoms \and X-rays \and Exospheres}
\end{abstract}

\section{Introduction}
\label{intro}
Traditionally, exospheric densities and velocity distributions
are modelled by spherical symmetric analytical Chamberlain functions
\citep{Chamberlain87}. 
Chamberlain theory assumes that gravity is the only force acting on the 
neutrals, that the exobase parameters (density and temperature) are 
uniform over a spherical exobase, and that no collisions occur above 
the exobase. 
Planetary exospheres are however not spherical symmetric due to non-uniform
exobase parameters and due to effects
such as photoionization, radiation pressure, charge exchange,
recombination and planetary rotation.
To account for these effects numerical simulations are needed.
Using Monte Carlo test particle simulations it is possible to
account for the above effects (if ion distributions are assumed).

Even though neutrals in the exospheres by definition do not
collide often, collisions occur.  Especially near the exobase
the transition is gradual from collision dominated regions
at lower heights (with Maxwellian velocity distributions) to
essentially collisionless regions at greater heights. 
Using test particles one can model collisions with an assumed 
background atmospheric profile~\citep{Hodges94}, but 
to account for collisions properly the test particle 
approach is not sufficient, and self consistent simulations are 
needed.  One approach to model collisions is the direct simulation 
Monte Carlo (DSMC) method~\citep{Bird} for rarefied flows, 
that has been applied to exospheres by~\cite{Shematovich05}. 

In this work we use the test particle approach to model the 
effects on the Martian exosphere from non-uniform exobase conditions, 
from photoionization, from radiation pressure, and from solar wind 
charge exchange. 
We launch test particles from the exobase and follow their trajectories. 
The forces on the particles are from gravity and radiation pressure. 
Along their trajectories the particles can be photoionized, and they can 
charge exchange with solar wind protons outside the bow shock. 
Exospheric column densities give us a qualitative estimate of how 
exospheric asymmetries effect solar wind charge exchange (SWCX) X-ray images. 
The energetic neutral atoms (ENAs) produced by charge exchange then gives 
us estimates of the ENA fluxes near Mars. 

In this work we do not include the effects of collisions since 
it greatly increases the computational cost. 
Collisions and photochemical reactions are two channels, in addition to
charge exchange, that produce an hot hydrogen population~\citep{Nagy90}. 
By only including charge exchange outside the bow shock as a 
source of hot hydrogen in this work we therefore under estimate the extent 
of the hydrogen corona.  Including all sources of hot hydrogen would 
result in more extended emissions of ENAs and X-rays, compared to 
the results presented here. 
An additional process that we do not consider is electron impact 
ionization, since it would require 
knowledge of electron fluxes and velocity distributions. 
However, this work should be seen as a first qualitative study 
of how asymmetries in exobase conditions at Mars effect the exosphere, 
and in turn the ENA and X-ray fluxes near Mars. 
To do a more accurate quantitative study is much more difficult. 
One then would need to specify the exact time, season and Mars--Sun 
distance; and have access to exobase conditions (at that time) from 
observations, global circulation models, and solar wind conditions, 
along with full knowledge of the ion fluxes near Mars. 

\subsection{ENA and X-ray imaging}
When the solar wind encounters a non-magnetized planet with an atmosphere, 
e.g., Mars or Venus, there will be a region of interaction, where solar wind 
ions collide with neutrals in the planet's exosphere.  Two of the processes 
taking place are
\begin{itemize}
\item The production of ENAs by charge-exchange 
between a solar wind proton and an exospheric neutral~\citep{JGR02}, and
\item The production of soft X-rays by SWCX 
      between heavy, highly charged, ions in the solar wind and exospheric 
      neutrals~\citep{GRL01}. 
\end{itemize}
Images of ENAs and SWCX X-rays can provide global, instantaneous, information 
on ion-fluxes and neutral densities in the interaction region.  It is however 
not easy to extract this information from the measured line of sight integrals
that are convolutions of the ion-fluxes and the neutral densities.  
We need to introduce models that reduce the complexity of the problem.  
At Mars, the hydrogen exosphere is enlarged due to the planet's low gravity, 
and thus provide a large interaction region, extending outward several planet 
radii. 
Traditionally, most of the modeling of the outer parts of Mars' exosphere has 
been using analytical, spherical symmetric, Chamberlain profiles.  
Planetary exospheres are however not spherical symmetric to any good 
approximation, and asymmetries at Mars observed in ENAs by Mars Express 
and in X-rays by XMM-Newton could 
be due to asymmetries in the exosphere.  
The neutral particle imager, part of ASPERA-4 on-board Mars Express, has 
observed asymmetries in the ENA fluxes in the shadow of the 
planet~\citep{Brinkfeldt06}.  The decline of ENA fluxes when entering 
the shadow is different from the rise in flux when exiting the shadow. 
The XMM Newton X-ray telescope observed Mars in November 2003, 
and SWCX X-rays were positively identified from the hydrogen 
corona~\citep{Dennerl06}. 
The morphology of the images are however different from what have 
been predicted by simulations~\citep{GRL01}.  One of the differences 
is that the emissions are asymmetric with respect to the ecliptic plane. 
Here we investigate the asymmetries in exospheric densities at Mars 
due to various factors, and their impact on ENA and SWCX X-ray images.

We may note that 
although asymmetric exospheres have not been used often in modeling of 
solar wind-Mars interactions, they are well known in the engineering 
community since aerobraking and satellite drag is directly dependent 
on exospheric densities, and provides total density 
measurements~\citep{MarsGRAM}. 

In Section~\ref{sec:1} we describe in more detail the methods and parameters 
used in our simulations.  In Section~\ref{sec:2} we then present the 
results of our numerical experiments, and finally we present conclusions 
in Section~\ref{sec:3}.

\section{Methods}
\label{sec:1}
Here we first describe the algorithms used to simulate Mars' hydrogen 
exosphere, and we then describe the detailed setup used in the numerical 
experiments. 

\subsection{The simulation algorithm}
\label{ssec:1}
In what follows, the coordinate system used is Mars solar ecliptic 
coordinates, centered at the planet with the $x$-axis toward the Sun, 
the $z$-axis perpendicular to the planet's velocity, 
in the northern ecliptic hemisphere, and a $y$-axis that completes 
the right handed system.  
Based on this solar ecliptic coordinate system we define (longitude, 
latitude) coordinates, with the $z$-axis toward 90 degree latitude, 
the $x$-axis (sub solar point) at (0,0), and the $y$-axis at (90,0). 

The simulation domain is bounded by two spherical shells centered at Mars. 
An inner boundary (the exobase) with a radius of 3580 km corresponding to 
a height of 200 km above the planet -- assuming from now on a planet radius 
$R_M$ of 3380~km --
and an outer boundary with a radius of 10~$R_M$. 
At the start of the simulation the domain is empty of particles. 
Then meta-particles are launched from the inner boundary at a rate of 
1000 meta-particles per second.  Each meta-particles corresponds 
to $N_m$ hydrogen atoms.  The location on the inner boundary of each launched 
particle is randomly drawn with probability proportional to the local 
hydrogen exobase density.  The velocity of each launched particle is 
randomly drawn from a probability distribution proportional to 
\[
  \left(\mathbf{n}\cdot\mathbf{v}\right) e^{-a|\mathbf{v}|^2}, 
\]
where $\mathbf{n}$ is the local unit surface normal, 
$\mathbf{v}$ is the velocity of the particle, and 
$a=m/(2kT)$, $m$ is the mass of a neutral, $k$ is Boltzmann's constant, 
and $T$ is the temperature (at the exobase position). 
Note that the distribution used is not a Maxwellian, but the distribution 
of the flux through a surface (the exobase) given a 
Maxwellian distribution at the location~\citep{Garcia}. 

After an hydrogen atom is launched from the inner boundary, we numerically 
integrate its trajectory with a time step of 5 seconds. 
To avoid energy dissipation, the time advance of the particles 
is done using a fourth order accurate symplectic integrator 
derived by~\cite{Candy91}. 

Between time steps, the following events can occur for an exospheric atom
\begin{itemize}
\item 
      Collision with an UV photon.  Following \cite{Hodges94} this 
      occurs as an absorption of the photon ($\Delta v$ opposite the 
      sun direction) followed by isotropic reradiation ($\Delta v$ in 
      a random direction).  
      From \cite{Hodges94} we use a velocity change $\Delta v=3.27$~m/s. 
      The collision rate used is $10^{-3}$~s$^{-1}$, 
      and the rate is zero if the particle is in the shadow behind the planet.
\item Charge exchange with a solar wind proton. 
      If the hydrogen atom is outside Mars' bow shock it can charge 
      exchange with a solar wind proton, producing an ENA, 
      at a rate of $8.4\cdot 10^{-8}$~s$^{-1}$. 
      The ENA is randomly drawn from a Maxwellian velocity distribution 
      with a bulk velocity of 450~km/s in the anti-sunward direction, and 
      a temperature of $1.2\cdot 10^5$~K.  
      Thus, the original exospheric hydrogen 
      atom is replaced by the ENA in the simulation. 
      Following \cite{Slavin91}, we define the bow shock by the surface 
      $(x,\rho)$~$R_M$ such that 
      \[
        x = \frac{-x_0+Le+x_0e^2-\sqrt{\rho^2e^2-\rho^2L^2}}{e^2-1}, 
      \]
      where $L=2.04$~$R_M$, $e=1.02$, and $x_0=0.55$~$R_m$. 
      Here $\rho=\sqrt{y^2+z^2}$ is the distance to the $x$-axis 
      (the Mars--Sun line). 
      We can note that the charge exchange rate gives an average 
      life time for an hydrogen atom of more than 100 days in the solar wind. 
      The implication for our simulations is that few ENA meta-particles are 
      produced.  To handle this we increase the charge exchange rate 
      by a factor of $f=1000$ and when a charge exchange event occurs, 
      the exospheric meta-particle with weight $N_m$ is replaced by a 
      meta-particle with weight $(1-1/f)N_m$ and an ENA with weight 
      $N_m/f$. 
\item Photoionization by a solar photon  
      occurs at a rate of $10^{-7}$~s when an exospheric 
      hydrogen atom is outside the optical shadow behind the planet, 
      and then the meta-particle is removed from the simulation. 
\end{itemize}
All rates above are from \cite{Hodges94} for Earth, and average 
solar conditions, scaled by~0.43 to account for the smaller fluxes at Mars. 
For a given event rate, $\tau$, after each time step, for each meta-particle, 
we draw a random time from an exponential distribution with mean  
$\tau$, and the event occur if this time is smaller than the time step.
Note that we only consider ENAs produced outside the bow shock, so 
the fluxes presented here is a lower bound.  Additional ENAs are 
produced inside the bow shock, but including those 
would require a complete ion flow model.  Anyhow, simulations \citep{JGR02} 
suggest that the ENA flux from the solar wind population is dominant in 
intensity. 
Also, we do not consider collisions between neutrals, as discussed 
in the introduction. We can note that omitting collisions means 
that the population of particles on satellite orbits will be small. 
The only generation mechanism for satellite particles will be 
radiation pressure.

\subsection{The simulation setup}
\label{ssec:2}
\begin{figure}
\centering
  \includegraphics[width=0.75\textwidth]{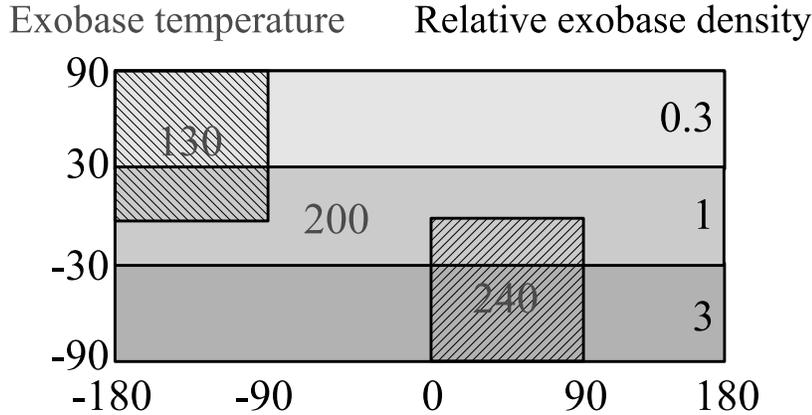}
\caption{
The non-uniform exobase density and temperature for Mars' hydrogen 
exosphere used in this study.  The coordinate system is Mars 
solar ecliptic longitude and latitude, with (0,0) corresponding 
to the sub solar point. The exobase temperature is shown in gray, 
with a value of 200~K, except for a region of increased temperature 
on the dayside, and a region of decreased temperature on the nightside. 
The exobase density is constant in three longitude bands, and 
shown in black is the density relative to a reference density 
of $4.2\cdot 10^5$~cm$^{-3}$. 
}
\label{exobase}
\end{figure}

As stated in the introduction, the aim of this study is to make a 
qualitative study of the effects of non-uniform exobase conditions 
on the hydrogen exosphere, and the implications for ENA and SWCX 
X-ray fluxes. 
Thus, we choose to study a simplified model problem where we have 
artificially chosen a spatial distribution of exobase density and 
temperature, shown in Figure~\ref{exobase}.   
We have constant density in three longitude bands, and three different 
temperature regions.  This is an approximation of the conditions 
at southern summer solstice, and was chosen as follows. 
We use the density and temperature for solar minimum conditions from 
\cite[Fig.\ 1]{Krasnopolsky02} at a height of 200~km; 
200~K and $4.2\cdot 10^5$~cm$^{-3}$ as a reference value. 
This is a day side average for a solar zenith angle of 60~degrees.
The corresponding density at 130~km (mostly CO$_2$) is 2.9 kg/km$^3$. 
Using the spatial variations from~\cite[Fig.\ 5 and 10]{Bougher00} 
we scale the reference values, and construct the exobase conditions 
shown in Figure~\ref{exobase}. 
We will later denote this the non-uniform case, and the case when 
we use the reference values for all of the exobase will be the 
uniform case. 
The spatial variations in \citep{Bougher00} are from a global circulation 
model of Mars' exosphere and is based on the observations available at 
that time.  
Later the model has been partially verified by observations~\citep{Lillis05}. 
Note that these exobase parameters specify the upward velocity distribution 
of neutrals at the inner boundary (the exobase).  The downward flux is 
then obtained from the simulation.  Therefore, these parameters will 
differ from the values obtained from the converged simulation, e.g., 
we can see in Figure~\ref{profiles} that the number density at the 
inner boundary has the proportions 1, 3, and 7 in the different 
latitude bands.

\section{Numerical experiments}
\label{sec:2}
First we investigate the effects of non-uniform exobase conditions 
on the hydrogen exosphere.  Then we study the implications for ENA and SWCX 
X-ray fluxes. 
\subsection{The exosphere}
\begin{figure}
\centering
  \includegraphics[width=0.75\textwidth]{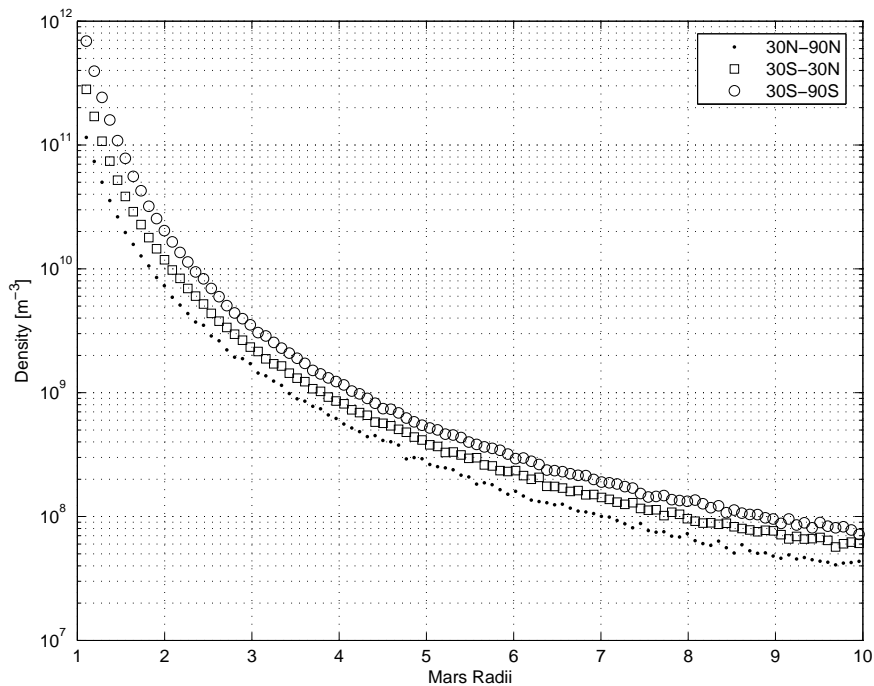}
\caption{Average hydrogen number density [m$^{-3}$] as a function of 
  planetocentric distance in~$R_M$ for non-uniform exobase conditions.  
  The exosphere has been divided into three regions corresponding to the three 
  latitude bands in Figure~\ref{exobase}, and each of the profiles shown 
  is an average over the corresponding region.  
  Latitude 30 to 90 degrees (dots), -30 to 30 (squares), 
  and -90 to -30 (circles). 
}
\label{profiles}
\end{figure}

Here we use the non-uniform exobase conditions shown in Figure~\ref{exobase}. 
First of all we want to examine how far out from the planet the 
exosphere is non-uniform.  Since the simulation particles are 
launched on ballistic trajectories, at any point there will be a mix of 
particles from different regions of the exobase.  This will 
introduce a smoothing of the exobase boundary conditions, and 
it is not obvious how large this smoothing will be, i.e.\ how far 
from the planet the non-uniformity will persist. 
To investigate this we divide the exosphere into three regions 
corresponding to the three latitude bands in Figure~\ref{exobase}, 
and plot profiles of the average hydrogen density for each of the 
regions.  These profiles at a time of 10~hours after the start of the 
simulation are shown in Figure~\ref{profiles}. 
We see that the density variation at the exobase by a factor of~10 
is reduced to approximately a factor of~3 at a planetocentric distance of 
2~$R_M$, and a factor of~2 at 10~$R_M$. 
Thus, the exospheric densities get more uniform with distance to the 
planet, but large differences in density persist all through the simulation 
domain. 

\subsection{SWCX X-rays}
\begin{figure}
\centering
  \includegraphics[width=0.75\textwidth]{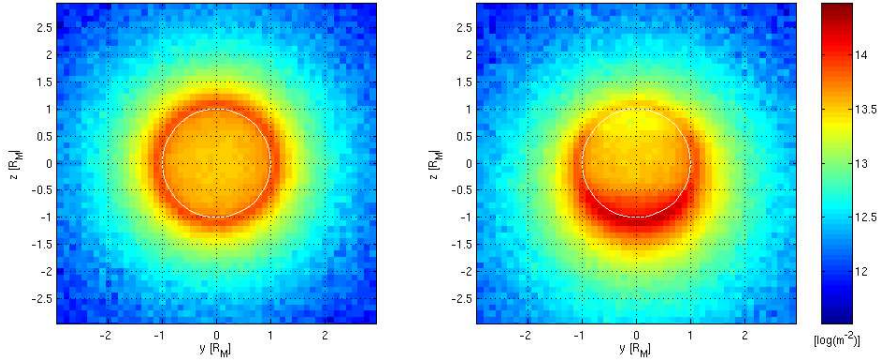}
\caption{Logarithm of the hydrogen column density [m$^{-2}$] 
along the $x$-axis for 
uniform (left) and non-uniform (right) exobase conditions. 
The white circles shows the size of the exobase. 
The time is 10~hours, the axes' units are $R_M$
and the total number of meta-particles are 848320 and 893240. 
The maximum column density is 0.781 and 1.86 
$\cdot 10^{14}$m$^{-2}$. 
}
\label{non-uni}
\end{figure}

To estimate the effects of the non-uniform exosphere on SWCX 
X-ray images, we note that the X-ray flux is a line-of-sight 
convolution of ion flux and hydrogen density.  
So in the unperturbed solar wind, outside the bow shock, the X-ray 
flux should be proportional to the hydrogen column density. 
In Figure~\ref{non-uni} we show the hydrogen column density for 
the cases of uniform and non-uniform exobase conditions. 
These are then estimates of what SWCX X-ray images would look like, 
from Earth at Mars' opposition, 
at least away from the planet (near the planet X-ray fluorescence 
dominate anyway).  
Note that the column density can vary by almost an order of magnitude, 
for constant planetocentric distances, even far away from the planet, 
in the non-uniform case. 
This would directly effect SWCX X-ray images and lead to 
asymmetries of the same magnitude. 

\subsection{ENA fluxes}
\begin{figure}
\centering
  \includegraphics[width=0.75\textwidth]{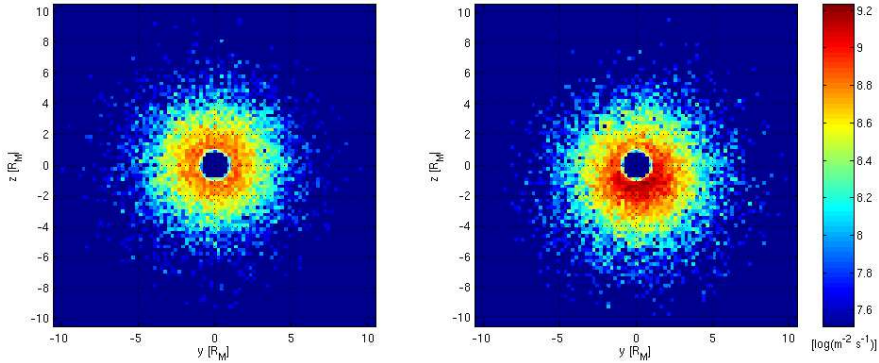}
\caption{Logarithm of the ENA fluxes [m$^{-2}$s$^{-1}$] through the 
$yz$-plane at $x=-1.0$~$R_M$ for uniform exobase conditions (left), 
and for non-uniform conditions (right). 
The maximum flux is $0.912$ and $1.79$ 
$\cdot 10^9$m$^{-2}$s$^{-1}$. 
The fluxes are computed by averages over all ENAs with $-1.05<x<-0.95$, 
from time 0 to 10~hours, the axes' units are $R_M$, 
and the total number of ENA meta-particles is 18796 and 22924. 
The white circle shows the size of the exobase. 
} \label{uni-ena}
\end{figure}

Here we investigate the fluxes of hydrogen ENAs that are created outside 
the bow shock by charge exchange between exospheric hydrogen and solar 
wind protons, with respect to any asymmetries induced by the 
asymmetric exosphere. 
One motivation for this investigation is that the neutral 
particle imager, part of ASPERA-4 on-board Mars Express, has 
seen asymmetries in the ENA fluxes in the tail behind the 
planet~\citep{Brinkfeldt06}. 
In Figure~\ref{uni-ena} we compare the ENA fluxes through a plane 
at $x=-1.0$ for uniform and non-uniform exobase conditions. 
We can note the spherical symmetry for the case of uniform 
exobase conditions, apart from the statistical fluctuations 
associated with test particle Monte Carlo simulations. 
On the other hand, in the case of non-uniform exobase parameters 
the asymmetry of the ENA flux is clearly visible as enhanced, 
and extended, flux in the south corresponding to the higher 
densities in the southern hemisphere. 
There is also a suggestion of enhanced densities in the 
$+y$ hemisphere corresponding to the enhanced exobase temperature 
in that hemisphere. 
For a constant planetocentric distance in this plane we see 
that the ENA flux can vary by more than a factor of~2. 

\begin{figure}
\centering
  \includegraphics[width=0.75\textwidth]{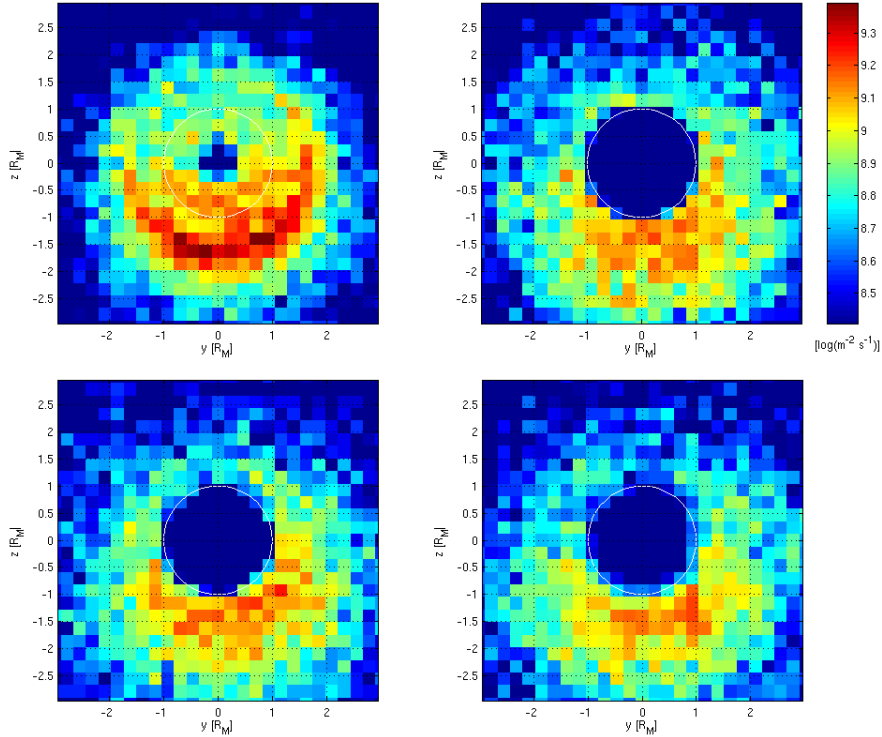}
\caption{Logarithm of the ENA fluxes [m$^{-2}$s$^{-1}$] through the 
$yz$-planes at $x=1$~$R_M$ (upper left), $x=0$ (upper right), 
$x=-1$ (lower left), and $x=-3$ (lower right) 
for non-uniform exobase conditions. 
All are averages over an $x$-width of 0.1~$R_M$. 
The maximum fluxes are 2.50, 1.63, 1.79, and 1.60 
$\cdot 10^9$m$^{-2}$s$^{-1}$. 
The white circle shows the size of the exobase. 
} \label{slices}
\end{figure}

How does the ENA fluxes vary at different positions relative 
to the planet? 
In Figure~\ref{slices} we plot the fluxes through the $yz$-planes 
at $x=$ 1.0, 0.0, -1.0, and -3.0. 
The north-south asymmetry is visible in all plots, with the highest, 
most concentrated fluxes at $x=1$. 
The area of large flux then spreads out slightly and has a bit lower 
intensity toward the tail. 
Note however that the flux at $x=-3$ is as large as the flux at 
$x=0$.  For all plots the maximum flux seems to be obtained at 
an approximate distance from the $x$-axis of 5000~km (about 1600~km 
outside the optical shadow).  
This is perhaps a bit surprising --- that the maximum flux is not 
closer to the umbra, but is a consequence of the shape of the 
bow shock in combination with the exospheric profiles, as seen in 
the flux through $x=1$ that is a crescent well outside the planet outline. 

In all numerical experiments above, radiation pressure and photoionization 
was not included.  It was found that including those events, as described 
in the previous section, did not change the results presented in 
any significant way. 

\section{Conclusions}
\label{sec:3}
Traditionally, modeling of the solar wind interaction with Mars' 
exosphere, and the production of SWCX X-rays and ENAs, has assumed 
a spherical symmetric exosphere.  From observations and simulations 
we know however that the exosphere is not symmetric.  
From the results of our simple test particle model of Mars' exosphere 
we find that asymmetries in exobase density and temperature propagate 
to large heights (many Martian radii). 
Column densities can deviate by almost an order of 
magnitude from symmetry, implying similar asymmetries in SWCX X-ray 
images. 
We also find that the fluxes of ENAs that are produced in the solar wind 
can deviate by more than a factor of two from symmetry. 
These asymmetries could explain the asymmetries seen in X-ray 
images and in ENA observations, but further studies are needed 
to find out if that is the case. 
We also find that radiation pressure and photoionization are 
unimportant processes in comparison to asymmetries in exobase parameters. 
Finally, we can note that 
asymmetries in the exosphere could also possibly explain the low exospheric 
densities seen by the neutral particle detector on-board Mars Express, as 
reported in this issue by \cite{Galli06}, since that measurement was over the 
northern hemisphere during early northern spring (April 25, 2004), when 
exospheric densities should have been low due to the seasonal variations.

\begin{acknowledgements}
Parts of this work was accomplished while the author visited NASA's 
Goddard Space Flight Center during 2005, funded by the 
National Research Council (NRC). The software used in this work was in 
part developed by the DOE-supported ASC / Alliance Center for Astrophysical 
Thermonuclear Flashes at the University of Chicago.
\end{acknowledgements}

\bibliographystyle{spbasic}
\bibliography{exospheres}   

\end{document}